\pdfoutput=1 
\documentclass[aps,pre,amsmath,amsymb,reprint,superscriptaddress]{revtex4-1}
\usepackage{graphicx}
\usepackage{bm}
\usepackage{xcolor}
\usepackage{relsize}
\usepackage{hyperref}
\usepackage{hycolor}
\hypersetup{breaklinks=true,colorlinks=true,citecolor=[rgb]{0,0.18,0.65}, 
urlcolor=[rgb]{0,0.18,0.65}, linkcolor=[rgb]{0,0.18,0.65}, final = true}

\def\eg{{\it{e.g.,~}}}
\def\ie{{\it{i.e.,~}}}
\def\kB{{k_{\mathrm{B}}}}
\def\DmuKPR{{\Delta\mu_\mathsmaller{\mathrm{KPR}}}}
\def\JKPR{{J_\mathsmaller{\mathrm{KPR}}}}
\def\QQ{{\mathcal{Q}}}

\begin{document}
	\title{Kinetic Proofreading and the Limits of Thermodynamic Uncertainty}
	\author{William D. Pi\~{n}eros} 
	\affiliation{Center for Soft and Living Matter, Institute for Basic Science (IBS), Ulsan 44919, Korea}
	\author{Tsvi Tlusty}
	\email{tsvitlusty@gmail.com}
	\affiliation{Center for Soft and Living Matter, Institute for Basic Science (IBS), Ulsan 44919, Korea}
	\affiliation{Department of Physics, Ulsan National Institute of Science and Technology (UNIST), Ulsan 44919, Korea}
		\date{\today}
	
	\begin{abstract}
	To mitigate errors induced by the cell's heterogeneous noisy environment, its main information channels and production networks utilize the kinetic proofreading (KPR) mechanism. Here, we examine two extensively-studied KPR circuits, DNA replication by the T7 DNA polymerase and translation by the E. coli ribosome. Using experimental data, we analyze the performance of these two vital systems in light of the fundamental bounds set by the recently-discovered thermodynamic uncertainty relation (TUR), which places an inherent trade-off between the precision of a desirable output and the amount of energy dissipation required. We show that the DNA polymerase operates close to the TUR lower bound, while the ribosome operates $\sim5$ times farther from this bound. This difference originates from the enhanced binding discrimination of the polymerase which allows it to operate effectively as a reduced reaction cycle prioritizing correct product formation. We show that approaching this limit also decouples the thermodynamic uncertainty factor from speed and error, thereby relaxing the accuracy-speed trade-off of the system. Altogether, our results show that operating near this reduced cycle limit not only minimizes thermodynamic uncertainty, but also results in global performance enhancement of KPR circuits.  	    	
 	\end{abstract} 

	\maketitle 

\section{Introduction} 
 	Fast and accurate processing of molecular information is essential for proper control, growth and regulation in the living cell. In carrying out essential tasks like protein synthesis, ribosomes are known to operate at error levels of $10^{-3}{-}10^{-4}$ incorrect peptide bindings per cycle~\cite{RibosomeErrEcoli,RibosomeTranscriptionErrRate,RibosomeErrors}, with even lower error rates for polymerases carrying out RNA transcription ($10^{-5}{-}10^{-6}$)~\cite{TranscriptionErrEukaryote,TranscriptionErrEcoli,TranscriptionErrEcoli2} and DNA replication ($10^{-7}{-}10^{-9}$)~\cite{PolymeraseErr,TranslationErrEColi,TranslationMutationRate}. 
 	This implies that the involved molecular machines must readily discriminate and preferentially bind the correct substrates over very similar, yet incorrect, competing substrates. However, simple energy discriminant binding models imply prohibitive energy binding differences among the pool of analogous substrates, and fail to predict the high level of fidelity observed. 
 	Instead, as originally proposed by Hopfield~\cite{HopfieldProofReading} and Ninio~\cite{NinioProofReading}, high accuracy may be achieved through kinetic proofreading (KPR), a mechanism that couples an effectively irreversible process to an intermediate reaction step which can then preferentially react or discard substrates --- via kinetic discrimination --- further down in the reaction pathway. In this manner, the original discrimination step that relies on binding energy differences is amplified through a second round of substrate verification. This mechanism has been confirmed experimentally for a variety of polymerase and ribosome systems~\cite{ProofReadingPolyRev,ProofReadingExperimental,ProofReadingExpRibosome}, and later recognized in signal transduction~\cite{KineticProofSignaling,KineticProofReadingSignalingRev} and homologous recombination~\cite{KPRRecombination}. 

	While KPR facilitates high fidelity synthesis, it imparts a significant energy cost to the overall process. Furthermore, the nanometric scale of these molecular systems renders them vulnerable to strong thermal and active fluctuations from the cellular environment, suggesting performance limits set by fundamental thermodynamic fluctuation-dissipation trade-offs~\cite{NonEqDynLivingSysRev,MolecularMachinesRev,ThermodynamicsNanoScale}. 
	Indeed, recent work on generic KPR models linked the amount of energy dissipated to the loss of configurational entropy during accurate product formation, and found that the efficiency of this process decreased rapidly for increasing levels of accuracy~\cite{KineticProofReadingEfficiencyTradeoff}. 
	More generally, accuracy of the copying process was found to be fundamentally tied by the amount of excess work dissipated by the system, with higher dissipation corresponding to higher accuracy, independent of underlying system topology~\cite{KineticProofReadingErrEntropy}.  

 	In analyzing KPR processes, one typically coarse-grains the full complex biochemical system into a discrete set of states connected by stochastic transitions approximated as a Markov process. 
 	However, even under these simplified dynamics, thermodynamics places an inherent energetic cost to the output precision of an observed quantity. 
	This seminal result has been dubbed the thermodynamic uncertainty relation (TUR), which is expressed in terms of the trade-off measure $\QQ$ as
\begin{equation}
 	\QQ \equiv \dot{Q} t \epsilon^2(t) \ge 2 \kB T~,
\label{eq:Q}
\end{equation} 
where $T$ is the temperature, $\kB$ is the Boltzmann constant, $t$ is time, $\dot{Q}$ is the energy dissipation rate, and $\epsilon^2(t) = \mathrm{Var}X /\langle X \rangle^2$ is the ratio of mean and variance of a current observable $X$~\cite{ThermUncertainty}. In short, Eq. \ref{eq:Q} implies that a reduction in the uncertainty of an observable must be accompanied by a matching increase in energy consumption. Optimal trade-off is achieved in the the limit of vanishing energy use (\ie equilibrium) and with normally-distributed heat dissipation~\cite{ThermUncGaussianHeat}. The TUR was first shown to hold in the limit of linear response, and later proven to hold for any Markov jump process in the short or long time limits~\cite{ThermUncertaintyLongtProof,ThermUncertaintyShorttProof}. More recently, this relation has been shown to follow for currents from a generalized fluctuation theorem framework~\cite{ThermUncFlucTheorem}.

	In the context of enzymatic kinetics, the TUR has been used  to infer performance boundaries in molecular systems such as motors~\cite{ThermUncFanoFactor,ThermUncMolMotors,StochThermRevMolMotorExp}.
	For instance, in a study by Hyeon and Hwang~\cite{KinesinThermUnc}, the transport efficiency of microtubule protein motors was defined in terms of $\QQ$ and analyzed using experimental data, showing that $\QQ$ is sub-optimized within physiological ATP fuel levels and cargo loadings. Notably, while the studied wild-type kinesin protein operates near a minimum in $\QQ$, the defective mutant was about three times less efficient and did not display a minimum.   

	Clearly, the TUR not only places fundamental constraints on system performance, but highlights the degree to which present day molecular systems may have accommodated to this limitation. While extensive kinetic analysis of copying-fidelity or proofreading mechanisms have been advanced in various contexts~ \cite{RibosomeRateAccTradeOff,KineticProofReadingGenerilizedMicroTubule,KineticProofReadingSensing2,ThermodynamicsMolecularCopying,HopfieldAsymtopicSpeedEntropyErrorLimit,KPRAssemblyRecA,KPRCascade}, direct TUR analysis of experimental KPR systems is surprisingly lacking. To this end, we consider a recent work by Banerjee, Kolomeisky and Igoshin~\cite{KineticProofBanerjee} on the KPR systems of the E. coli ribosome and the T7 bacteriophage polymerase, and analyze these circuits in the context of TUR. In~\cite{KineticProofBanerjee}, the reaction networks were adapted to a standard Hopfield-Ninio KPR model using experimental values of the kinetic rate constants. Taking the physiological state as a reference point, they investigated the speed-accuracy trade-off as a function of the kinetic rates, finding that speed is prioritized over error rate. In a follow-up work on the same systems, they found that speed is also prioritized over energy dissipation and output noise~ \cite{KineticProofMallory}. 

	In this paper, we offer a complementary view on the existing body of analysis, focusing on the fundamental implications of the TUR on the \emph{synthesis process} in the KPR networks of E. coli ribosome and T7 DNA polymerase. We show that, in general, decreasing error rates and mean production times coincide with an underlying effective \emph{reduced network} of reactions steps that minimizes the TUR measure $\QQ$ of production. Approaching this reduced network not only provides the best energetic trade-off between production precision and energy dissipation through $\QQ$, but also decouples the operational speed from the dispersion of production. As a result, this regime minimizes trade-off constraints between the mean production time and the error rate for a given set of control parameters and fixed energy budget. Together, we show that approaching the reduced network regime corresponds to enhanced global performance of the studied ribosome and polymerase systems.  


\section{Kinetic Proofreading Circuits} 
\label{sec:Methods}
	\begin{table*}[!tbh]
	\caption{Kinetic model parameters as originally reported by Banerjee et al \cite{KineticProofBanerjee}. Kinetic rate constants $k^{(-)}_{i,R}$ reported in $s^{-1}$ and discrimination factors $f_{i}$ are dimensionless by definition. Rate constants $k^-_3$ and $k^-_p$ and discrimination factors $f^-_3$ and $f^-_p$ and are derived from the constraint relations of eq. \ref{eq:kconstraints} and \ref{eq:fconstraint}, respectively. 
	} 
	\label{tab:parameters}
	\begin{ruledtabular} 
	\begin{tabular}{lcccc}
	 Parameters	& Ribosome WT 		& Ribosome Acc 		& Ribosome Err 		& T7 polymerase  \\ \hline
	 $k_{1,R}$	& $40 $ 		& $27$ 			& $37$			& $250$	\\
	 $k^-_{1,R}$	& $0.5$ 		& $0.41$		& $0.43$		& $1$	\\
	 $k_{2,R}$	& $25 $ 		& $14$			& $31$			& $0.2$	\\
	 $k^-_{2,R}$	& $10^{-3}$  		& $10^{-3}$ 		& $10^{-3}$ 		& $700$	\\
	 $k_{3,R}$	& $8.5\times10^{-2}$  	& $4.8\times 10^{-2}$ 	& $7.7\times 10^{-2}$	& $900$	\\
	 $k_{p,R}$	& $8.415$ 		& $4.752$		& $7.623$		& $250$ \\
	 $f_{1}$	& $0.675$ 		& $0.926$		& $0.973$		& $8\times10^{-6}$\\
	 $f^-_{1}$	& $94$	 		& $112.2$		& $9.3$			& $1\times10^{-4}\footnote{Value chosen from same experimental limits to ensure positive values of $J_{pW}$.}$\\
	 $f_{2}$	& $4.8\times10^{-2} $ 	& $3.5\times10^{-2}$	& $0.126$		& $11.5$\\
	 $f^-_{2}$	& $1$ 			& $1$			& $1$ 			& $1$\\
	 $f_{3}$	& $7.9$ 		& $10.34$		& $7.65$		& $1$\\
	 $f_{p}$	& $4.2\times10^{-3}$ 	& $7.4\times10^{-4}$	& $4.1\times10^{-3}$	& $4.8\times10^{-5}$\\
	\end{tabular}
	\end{ruledtabular}
	\end{table*}

	\begin{figure}[ht!]
	\includegraphics[scale=0.20]{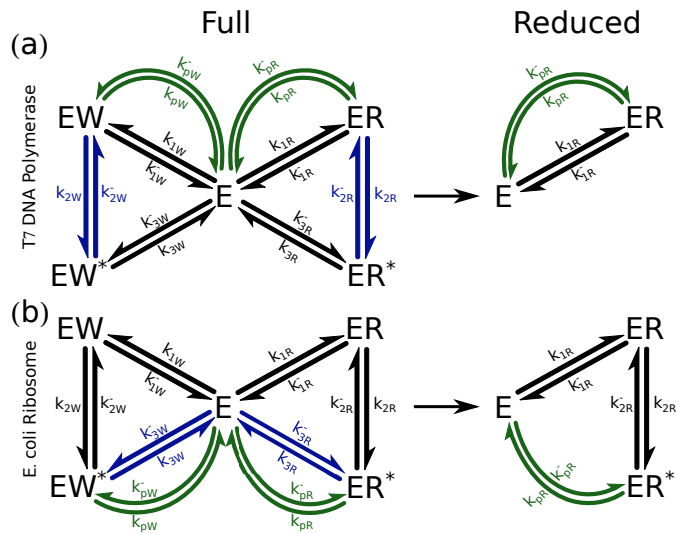}  
	\caption{Chemical reaction networks for (a) T7 DNA polymerase and (b) E. coli ribosome. Half arrows indicate reversible forward and backward paths for both the right (R) and wrong (W) cycles. Kinetic rates are labeled by $k^{(-)}_{i,W/R}$ where $i={1,2,3,p}$ for each relevant path. Note that transitions through $p$ in green (curved gray) half-arrows are implied to reset the enzyme to its initial state after product formation. Blue half-arrows (straight gray) indicate the proofreading transitions. Shown to the right of the black arrows are the reduced kinetic cycles (RC) where only relevant paths and rates leading to correct product formation are included.}
	\label{fig:networks} 
	\end{figure} 

	We study the standard Hopfield-Ninio KPR model as adapted to the T7 DNA polymerase and the E.coli ribosome by Banerjee et al.~\cite{KineticProofBanerjee}, using measured values of the kinetic rate constants. These networks, shown in Figure \ref{fig:networks}, capture the main steps of nucleotide ligation in the polymerase and peptide elongation in the ribosome. 

	In particular, these kinetic pathways model the steady-state processive action of the polymerase or ribosome and consist of the following three major steps. For the T7 DNA polymerase, the cycle starts at the polymerase bound to the growing DNA strand in state \textbf{E} where it adds either the correct (R) or incorrect (W) deoxy-NTP molecule to the growing strand and forms the \textbf{ER/W} state. The system can then ligate another dNTP molecule (path p) to restart the cycle or shift the strand to the polymerase exo-site \textbf{ER*/W*} where the dNTP is hydrolized and removed back to state \textbf{E}. Similarly for the ribosome cycle in (b), state \textbf{E} represents the mRNA template bound in the growing ribosome poly-peptide complex. Binding of the cognate (R) or non-cognate (W) aminoacyl-tRNA, EF-TU elongation factor and GTP leads to the second state \textbf{ER/W}. Hydrolysis of the GTP molecule brings the complex to state \textbf{ER*/W*} where the amino-acid can be added to the growing polypeptide strand (link p, green) or discarded (link 3,blue) which restarts the cycle. Note that the main difference in the topology of the cycles is that the KPR and production steps follow the first intermediate \textbf{ER/W} in the polymerase, whereas in the ribosome these occur only following the second hydrolyzed intermediate \textbf{ER*/W*}.

    All underlying kinetic rates $k^{(-)}_{i,W/R}$ for these cycles are assumed to be reversible and first order (\ie measured in $s^{-1}$ units) in constant substrate concentration at physiological conditions. Additionally, we maintain constant chemical potential differences $\Delta\mu$ of the underlying chemical reactions in both R and W cycles. This requirement constrains the rates as 
\begin{equation}
	\prod_{i=1}^{N} \frac{k_{i,W/R}}{k^-_{i,W/R}} = e^{\Delta\mu}~,
\label{eq:kconstraints}
\end{equation} 
where $\Delta\mu= \DmuKPR$ for proofreading cycles ($N=3$) or $\Delta\mu_p$ for production cycle ($N=p$) and chemical potentials are hereafter measured in $\kB T$ units. We take the approximate physiological values of the chemical potential differences, $\Delta\mu_{p} \sim 26 \kB T$ for poly-peptide elongation, $\Delta\mu_p \sim 11 \kB T$ for nucleotide ligation and $\DmuKPR \sim 20 \kB T$ for the hydrolysis KPR step in both systems~\cite{KineticProofMallory}. 
For convenience, we also define discrimination factors $f_i$  which relate the R and rate constants as $f^{(-)}_i=k^{(-)}_{i,W}/k^{(-)}_{i,R}$ and represent the biased discriminant enzyme behavior when bound to either the right or wrong substrate. 
These factors are similarly constrained through Eq. \ref{eq:kconstraints}, 
\begin{equation}
\label{eq:fconstraint}
	\prod_{i=1}^{N} \frac{f_i}{f^{-}_i} = 1	~.
\end{equation}  
Values for all $k_{i,R/W}^{(-)}$ and $f^{(-)}_i$ are as adapted by Banerjee et al.~\cite{KineticProofBanerjee} from experimental sources~\cite{KineticRiboExperiment,KineticPoly1,KineticPoly2,KineticPoly3} and listed in Table \ref{tab:parameters} above.   

	For the purpose of our thermodynamic uncertainty analysis, we calculate the TUR measure $\QQ$ for \emph{correct product transitions} across the path $p$ in the ribosome and polymerase as follows. The relative uncertainty $\epsilon^2(t)$ across this production path can be found from the mean transition current $J_{p}^{R}= \langle X \rangle /t$ and its diffusion constant $D_{p}^{R}=\mathrm{Var}X/2t$ so that
\begin{equation}
\label{eq:epsilon}
	\epsilon^2(t) = \mathrm{Var}X/\langle X \rangle^2  = 2 D_{p}^{R} / (J_{p}^{R})^2 t~.  
\end{equation} 
Using the definition for $\QQ$ in Eq.\ref{eq:Q} and $\epsilon$ above we thus get
\begin{equation}
	\mathcal{Q} = 2 \dot{Q} D_{p}^{R}/(J_{p}^{R})^2~, 
\end{equation} 
where the energy dissipation $\dot{Q} = \kB T \sigma$ is defined in terms of the entropy production rate $\sigma$ as 
\begin{equation}
\label{eq:sigma}
	\sigma = \JKPR \DmuKPR + J_{p} \Delta\mu_p~.
\end{equation}
The currents that determine $\sigma$ in Eq. \ref{eq:sigma} are $J_{p}=J_{p}^{R}+J_{p}^{W}$, the production current for both R and W cycles, and $\JKPR=J_{i}^{R}+J_{i}^{W}$, the discarded substrate current from kinetic proofreading (for $i=2$ in polymerase or $i=3$ ribosome). We also calculate the mean production time $\tau$ defined as the average time required to observe one net product addition onto the growing strand. This is given from the production rate as 
\begin{equation}
	\tau \equiv 1/J_{p}^{R}~. 
\end{equation}
This definition of time is equivalent to a mean passage time in the limit of irreversible product formation and vanishing incorrect product rate $k_{pW/R}$.  
 Similarly, the error $\eta$ is defined as the fraction of incorrect substrate units added onto a growing peptide chain or DNA strand,
\begin{equation}
	\eta = \frac{J_{p}^{W}}{J_{p}^{W}+J_{p}^{R}}~,
\end{equation}
where $J_{p}^{W}$ is the current across the $p$ link in the wrong W cycle. The values of the currents, $J_{p}^{R}$ and $J_{p}^{W}$, and the diffusion constant $D_{p}^{R}$, are calculated using Koza's steady-state method~\cite{KozaMethod}, as demonstrated extensively in other studies~\cite{ThermUncertainty,KineticProofMallory,ThermUncFanoFactor}.

    In addition to the full reaction networks, it is instructive to consider idealized cycles consisting of only states and paths leading to correct product formation (shown on the right side in Figure \ref{fig:networks}). These reduced cycles (RC) represent, by construction, perfect performance of the underlying protein systems. The RC circuits allow direct comparison of current-dependent metrics like $\QQ$ and $\tau$ between the actual and ideal system.
    Considering that $k_i \gg k^-_i$ for the systems studied here, the production current $J_{p}^{R}$ in these idealized cycles is a simplified function of the forward rate constants $k_i$ of the form $J_{p}^{R}\sim\prod_i^n k_i$.
    One can thus define the forward rate constants in terms of a single control parameter $k$ as $k_i \equiv a_i k$ where $a_i \equiv k^{\mathrm{phys}}_{i}/k^{\mathrm{phys}}_{1}$ are the ratios of the rate constants at physiological values. It follows that current in the idealized circuit $J_{p}^{R} \sim k^n \prod_i^n a_i$ is a monotonic function (a power) of $k$ that conserves the rate constant proportionality of the original system. The physiological state is matched when $k=k^{\mathrm{phys}}_{1}$. As we later show, operating near the regime of the ideal RC cycles implies minimimal $\QQ$ values, and affords enhanced accuracy/speed trade-off performance.

\section{Results and Discussion} 
\label{sec:Results} 
\begin{table}
\caption{Values of calculated TUR measure $\QQ$ (Eq. \ref{eq:Q}) and score ratios to its lower bound $\QQ_\mathrm{lh}$ for a given number of states $N$ and constant $\Delta\mu$ as defined in Eq. \ref{eq:Qbound}. For ribosomes, $N=3$ and $\Delta\mu=\Delta\mu_p=26 \kB T$. For T7 polymerase $N=2$ and $\Delta\mu=\Delta\mu_p=11 \kB T$.}
\label{tab:Qscores}
\begin{ruledtabular} 
\begin{tabular}{lcc}
	& $\QQ$ ($\kB T$)	& $\QQ/\QQ_\mathrm{lh}$ \\ \hline
Err Ribosome	& 137 	& 16  \\
WT Ribosome 	& 48   	& 5.6 \\
Acc Ribosome	& 28	& 3.2 \\
T7 Polymerase	& 7.1	& 1.3 \\
\end{tabular}
\end{ruledtabular}
\end{table}

	\begin{figure*}[htbp!]
	\includegraphics[scale=0.26]{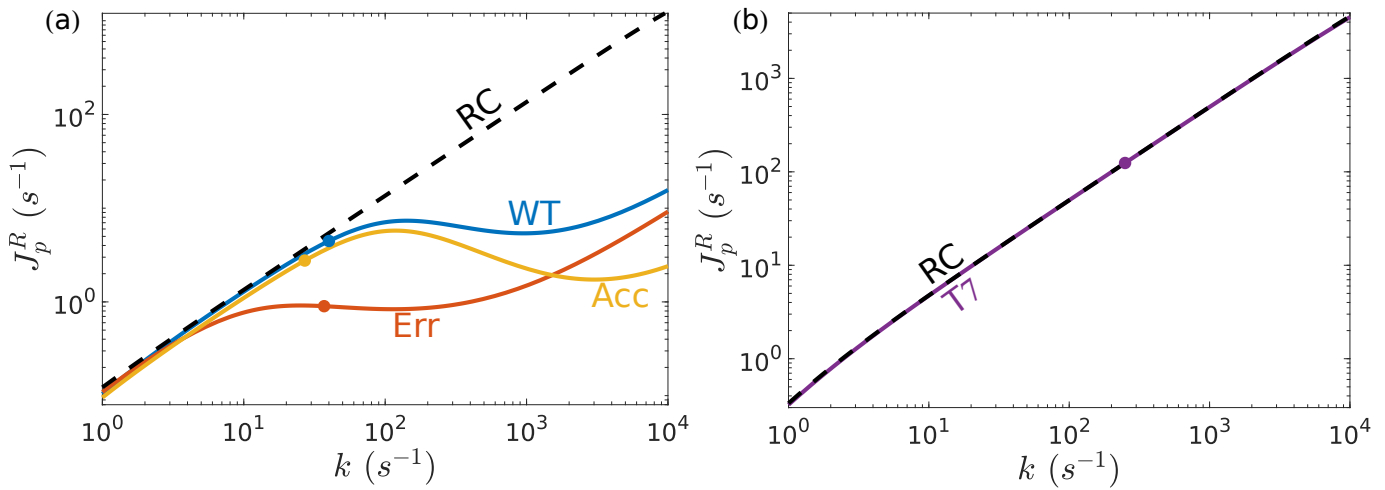} 
	\caption{Product cycle current $J_{p}^{R}$ as a function of generalized rate constant $k$, defined as $k_{1R}=a_1 k$, $k_{2R}=a_2 k$, $k_{pR}=a_p k$ with $a_1=1$, $a_2=k_{2R}^{\mathrm{phys}}/k_{1R}^{\mathrm{phys}}$ and $a_p=k_{p}^{\mathrm{phys}}/k_{1R}^{\mathrm{phys}}$, where $k_{i}^{\mathrm{phys}}$ are the physiological values. The physiological points are shown as dots. (a) The current of correct substrate production $J_{p}^{R}$ for Wild-type ribosome (WT, blue top), more erroneous (Err, red bottom) and more accurate (Acc, yellow middle) mutants and the ideal RC current (dashed line). (b) Same as (a) but for T7 polymerase (solid line). The difference between the actual T7 current and RC is in the order of $1\%$ and not noticeable at this scale.} 
	\label{fig:currents}
	\end{figure*} 
	
	\emph{The approach of KPR circuits to the TUR limit.---} We begin by reporting the physiological values of $\QQ$ for product transitions in the ribosome and polymerase systems as shown in  Table \ref{tab:Qscores}. Here, the T7 polymerase achieves the lowest value of $\QQ$ which is about seven times smaller than the native WT ribosome, with the more accurate mutant Acc closer to the limit than either WT or the less accurate Err.
	In order to compare these results more meaningfully, however, we must account for the underlying energy cost of cycle operation, which differs between the T7 DNAP and ribosome systems. 

	This can be achieved by considering the reduced cycles (RCs) consisting of only states and transitions leading to correct product formation  as introduced in section \ref{sec:Methods} and Figure \ref{fig:networks}(b). These RCs were extracted from the full network of states and represent an idealized limit where only the correct substrate is processed in the absence of any competing paths. Operating at the RC limit therefore provides the best overall enzyme performance for a fixed energy budget. In particular, the RC limit implies a unicycle regime for which a lower bound for $\QQ$ is known to be 
\begin{equation}
	\QQ \ge \QQ_\mathrm{lh} \equiv
	2 \kB T\left( \frac{\Delta\mu}{2 N} \coth \frac{\Delta\mu}{2 N}  \right ) \label{eq:Qbound}
\end{equation}
where $N$ denotes the number of states in the network, and $\Delta\mu$  is the overall change in Gibbs free energy of the underlying chemical reactions per cycle (in $\kB T$ units)~\cite{ThermUncMultiCycles,StochThermRevMolMotorExp}. This hyperbolic lower bound $\QQ_\mathrm{lh}$ is achieved for a system with uniform forward and backward rate constants and reduces to the minimal value of $2 \kB T$ in the vanishing $\Delta\mu$ limit. Thus, the hyperbolic bound is pertinent for far-from-equilibrium driven process such as KPR and represents the best  efficiency attainable given an energy input. The ratio of the physiological values of $\QQ/\QQ_\mathrm{lh}$ shown in Table \ref{tab:Qscores} are therefore a normalized optimization score, for a specific energetic constraint, of either the polymerase or ribosome systems. 
Markedly, the polymerase operates close to the TUR limit at $\QQ/\QQ_\mathrm{lh} \simeq 1.3$, while the ribosomes are $3-16$ times further away, even after accounting for the specific energy cost of the underlying chemical transitions. 

One can obtain a deeper appreciation of the score ratios $\QQ/\QQ_\mathrm{lh}$ by comparing the full enzymatic cycles to their respective RC limits. To achieve this, we define a collective rate constant $k$ which governs the product output current $J_{p}^{R} \sim k^n$ in RC networks, and from which $\QQ$ and other performance metrics are calculated as detailed in section \ref{sec:Methods}. Consequently, $k$ serves as a control parameter that allows for direct performance comparison between actual and idealized RC systems. Figure \ref{fig:currents},  presents $J_{p}^{R}$ as a function of $k$ for both RCs and the full ribosome and polymerase systems. As seen, the ideal current is increasing montonically with $k$, while the actual current is non-monotonic for the ribosome systems and nearly indistinguishable from RC for the T7 polymerase. These results are a first indication that the polymerase is indeed working at virtually reduced network conditions, hence the lower $\QQ$ value, while the ribosomes only approach this limit at longer operation times (lower currents). 
	
	\begin{figure*}[htbp!]
	\includegraphics[scale=0.23]{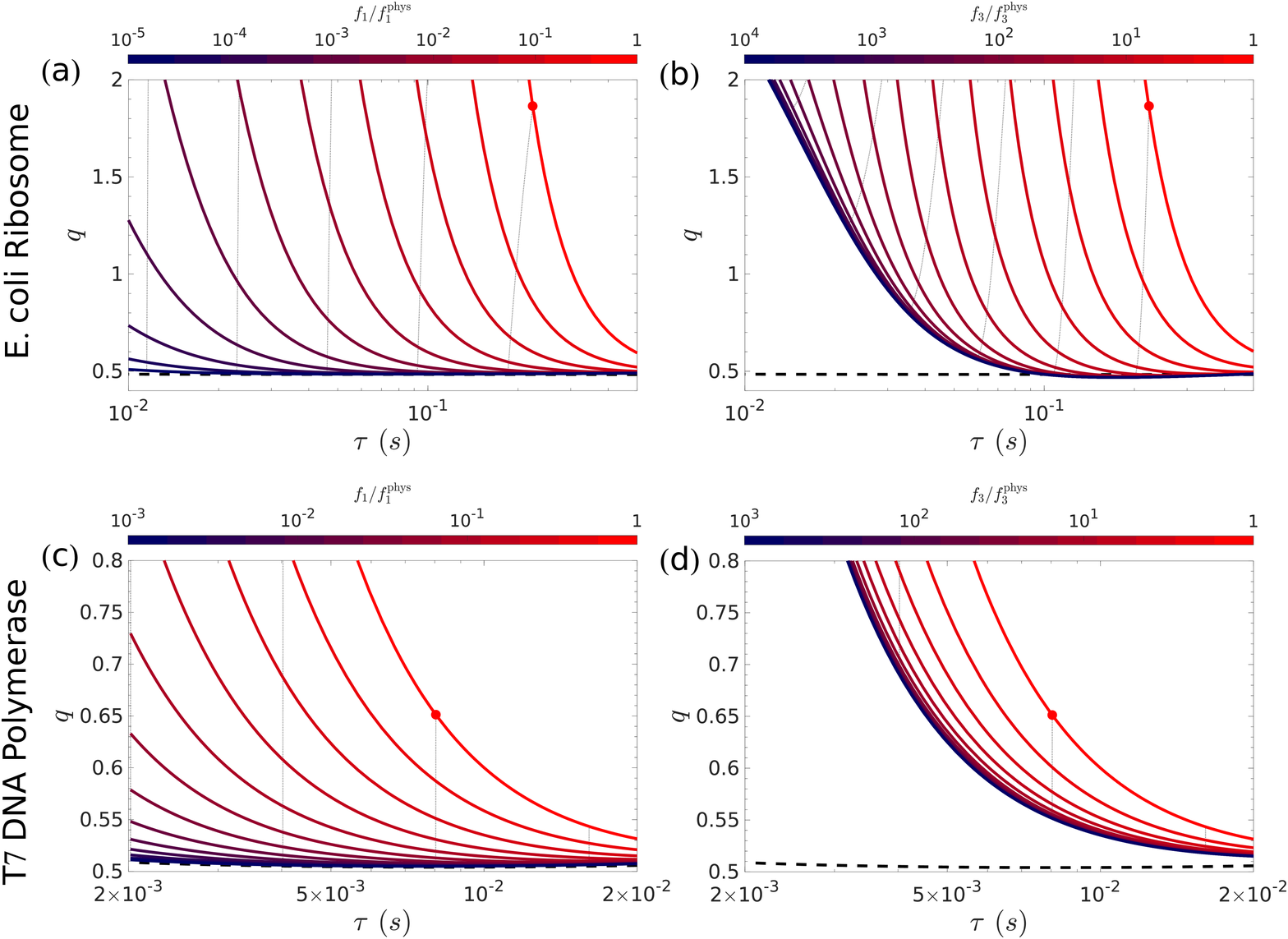} 
	\caption{Parametric plots of the normalized TUR measure $q \equiv \QQ/\Delta\mu_p$ vs. mean production time $\tau$ ($s$) as a function of $k$ for decreasing values of discrimination factor $f_1$ (a,c) and increasing values of the proofreading discrimination factor $f_3$ (b,d) for the wild-type ribosome and DNA polymerase respectively.
	Each corresponding factor $f_i$ is scaled from its physiological value $f^{\mathrm{phys}}_{i}$ as shown in the gradient scale on top. 
	Dashed black lines indicate ideal RC limit. Dotted gray lines indicate curves of constant $k$ scaled from its physiological value in powers of $2$.}
	\label{fig:QvTfs}
	\end{figure*} 
	
	The proximity of these systems to their lower TUR limits motivates a further examination of the full circuits and their corresponding RC limits. To this end, we consider two operating cases that either reduce or preserve the full reaction network respectively: (a) perfect binding discrimination corresponding to $f_1 \to 0 $ and b) perfect proofreading discrimination corresponding to $f_3 \to \infty$. 
	In case (a) the full system gradually reduces to the RC limit by entire omission of the $W$ branch. In contrast, case (b) preserves the overall system topology while minimizing the impact of incorrect synthesis in the W branch. As shown below, comparing $\QQ$, the error rate $\eta$ and the mean production time $\tau$ for either case allows us to see how the approach to the RC limit governs the performance of the full systems.
	
\emph{Approaching the RC limit decouples the TUR measure $\QQ$ from the mean production time $\tau$.---} 
To allow for normalized comparison of the actual system over its RC cycle, independent of system topology and energetic cost, we define the normalized TUR measure $q \equiv \QQ/\Delta\mu_p$. 
Figure \ref{fig:QvTfs} shows $q$ for the WT ribosome against the mean production time $\tau$, as a function of $k$ while $f_1$ (case (a)) or $f_3$ (case (b)) are parametrically varied. 
 As seen in (a) the WT ribosome displays a clear trade-off between $q$ and $\tau$ (red line), but quickly attenuates and decouples as it approaches the RC limit (black dashed line). On the other hand, increasing $f_3$ (b) maintains the trade-off constraint between $q$ and $\tau$ at all points, even when approaching the RC limit. Similar trends are seen for the polymerase in figure \ref{fig:QvTfs} (c) and (d). From these results we find that a system operating near the RC limit may more readily minimize both the product output noise and mean production time without being constrained by a strong trade-off relation. 
	\begin{figure*}[htbp!]
	\includegraphics[scale=0.27]{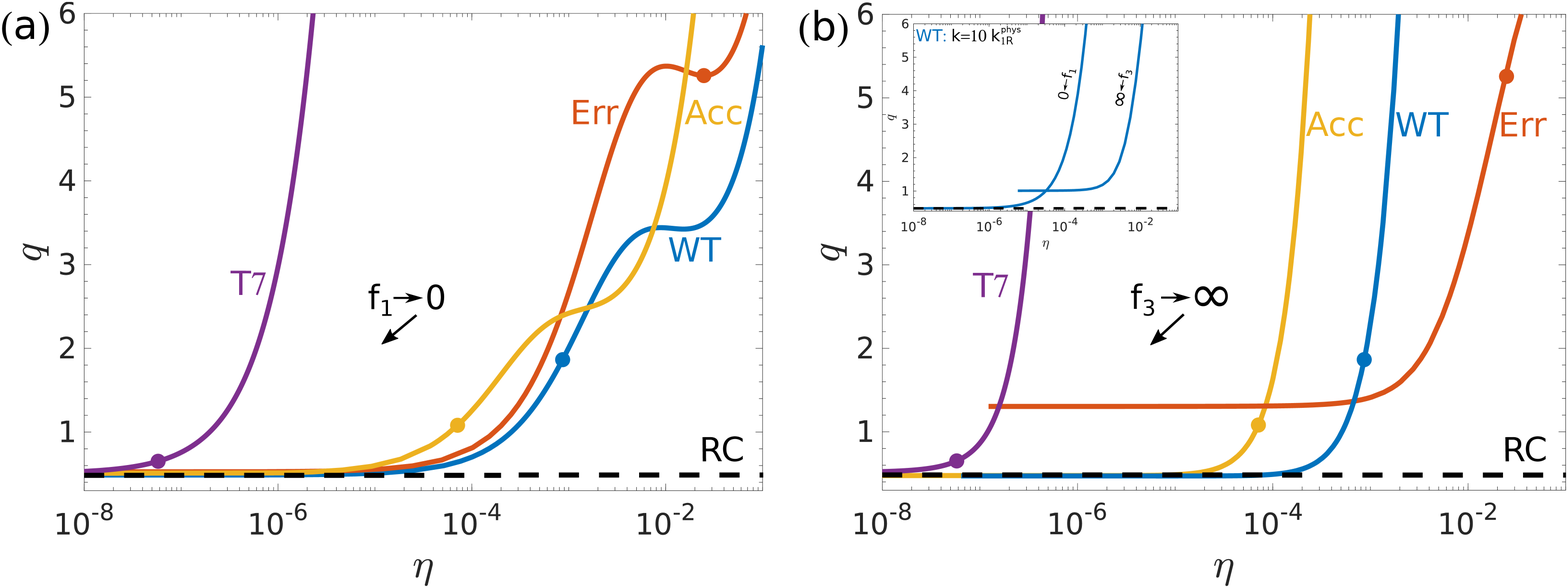} 
	\caption{a) Parametric plot of $q \equiv \QQ/\Delta\mu_p$ vs. error $\eta$ as a function of $f_1$. Lines indicate T7 DNA polymerase (purple leftmost), WT ribosome (blue bottom), erroneous (red top), and more accurate (yellow middle) ribosome mutants. Thick points indicate physiological values. Dashed line is a guide to the eye showing the value of $\QQ$ achieved at the ideal RC limit of WT ribosome which is approximately the same for all systems shown. (b) Same as (a) but for $f_3 \to \infty$. Inset: WT ribosome scaled from $k=k_{1R}^{\mathrm{phys}}$ to $k=10 k_{1R}^{\mathrm{phys}}$ illustrating that only $f_1 \to 0$ guarantees $\QQ$ goes to the RC limit.} 
	\label{fig:QvErrf1}
	\end{figure*} 

	\emph{The error rate $\eta$ decouples from the TUR measure $\QQ$ in high fidelity regimes.---} 
	Figure \ref{fig:QvErrf1} shows $q \equiv \QQ/\Delta\mu_p$ against the error $\eta$ for decreasing $f_1$ (a) and increasing $f_3$ (b) from the measured physiological values. In both cases, $q$ decreases with decreasing $\eta$ and becomes decoupled in the low error regime. However, this asymptotic value of $\QQ$ only matches the RC limit in the vanishing $f_1$ case (a) but not in case (b) where the Err mutant stays well above the RC value at the $f_3 \to \infty$ limit. The inset illustrates this asymptotic behavior more clearly where in this case $k$ has been rescaled from the physiological values as $k=10 k_{1}^{\mathrm{phys}}$ for the WT ribosome, and shows that the RC limit can be reached for $f_1\to 0$, but not for $f_3\to\infty$.  

	\begin{figure*}[ht]
	\includegraphics[scale=0.23]{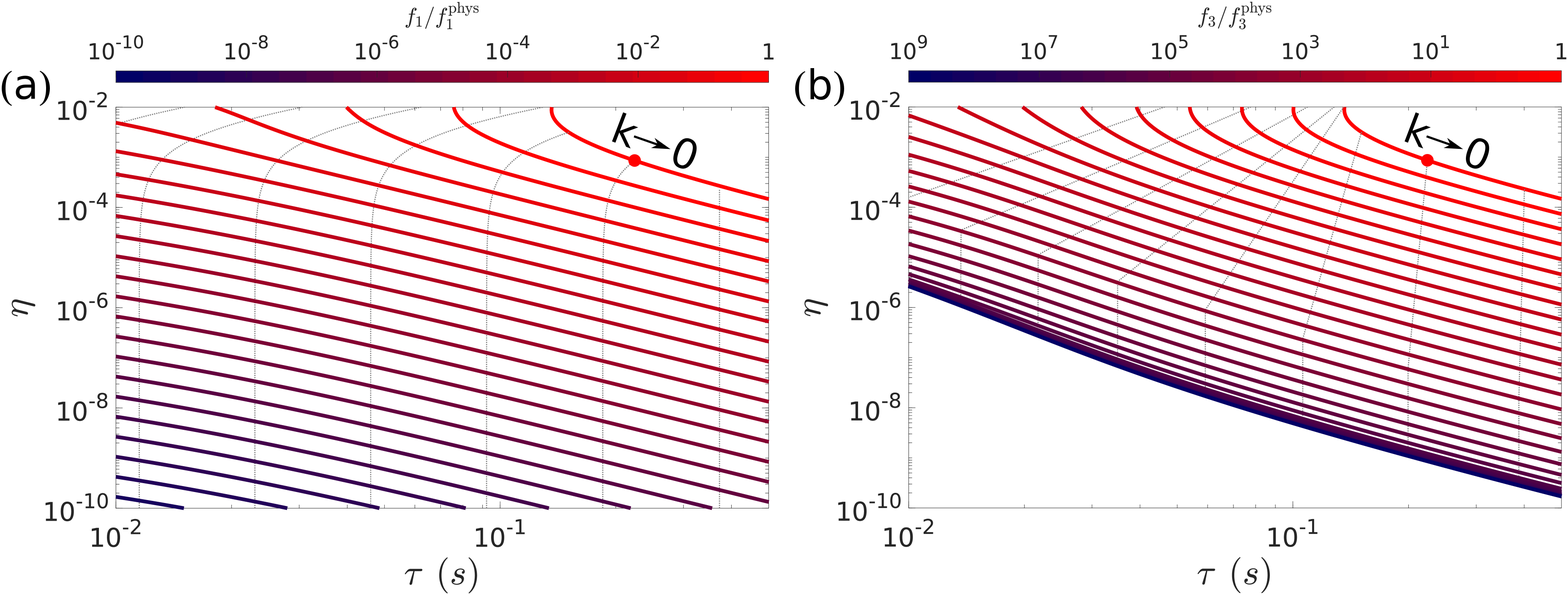}
	\caption{Parametric plots of the error $\eta$ versus mean production time $\tau$ for wild-type ribosome as a function of $k$ for (a) decreasing values of binding discrimination $f_1$, and (b) increasing values of proofreading discrimination $f_3$. Each respective factor $f_i$ is scaled from its physiological value $f^{\mathrm{phys}}_{i}$ as shown in the gradient scale on top. Dotted gray lines indicate curves of constant $k$ value scaled from its physiological value in powers of $2$.} 
	\label{fig:ErrvsTau} 
	\end{figure*} 
	
	\emph{Approaching the RC limit relaxes the trade-off constraint between error rate and mean production time.---} 
	It is also instructive to compare the error $\eta$ to the mean production time $\tau$ as a function of $k$ in the context of the idealized RC limit (Figure \ref{fig:ErrvsTau}). While changing either $f_1$ or $f_3$ parametrically is not expected to decouple the trade-off between measures, these curves highlight improved performance close to the RC limit in addition of minimizing $\QQ$. For instance, while increasing discriminant proofreading $f_3$ naturally improves the accuracy of the system, it ultimately approaches a best trade-off curve for this parameter variation (a Pareto front). In contrast, reducing $f_1$ weakens the trade-off relationship (smaller negative derivatives) while moving these trade-off curves arbitrarily close to the origin by construction (incorrect substrate is never bound).  

	\begin{figure}[htbp!]
	\includegraphics[scale=0.25]{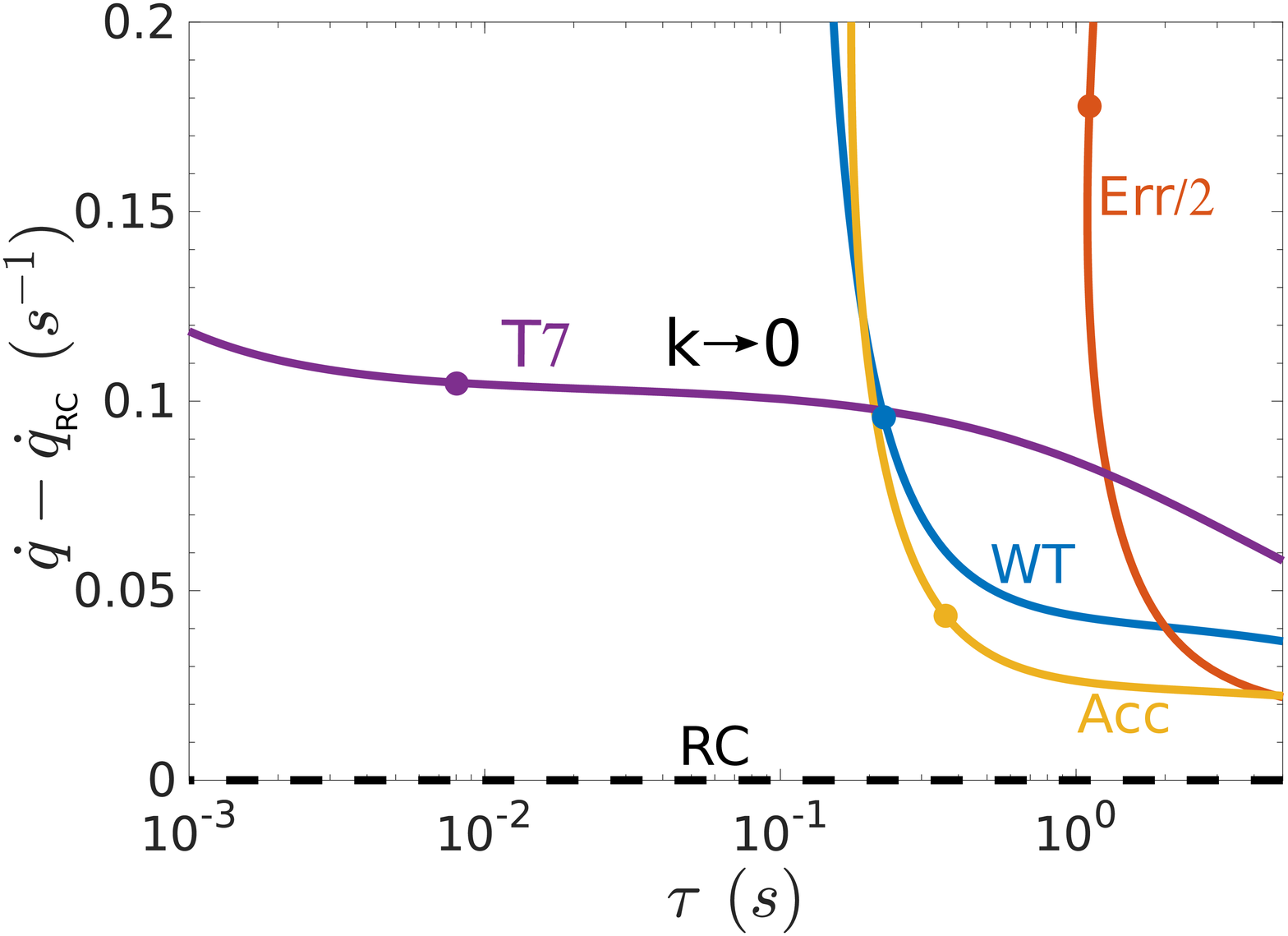} 
	\caption{Parametric plots of normalized dissipation differences between the actual $\dot{q}=\dot{Q}/\Delta\mu_p$ and the ideal RC limit $\dot{q}_{\mathsmaller{\mathrm{RC}}}=\dot{Q}_\mathsmaller{\mathrm{RC}}/\Delta\mu_p$ versus the mean production time $\tau$ as a function of $k$. Lines indicate T7 DNA polymerase (purple left), WT ribosome (blue middle), erroneous (red top), and more accurate (yellow bottom) ribosome mutants. Err mutant plot has been scaled down by a factor of $2$ to fit the figure. Points indicate physiological values for each line respectively. Dashed line marks the RC difference which is zero by definition.}
	\label{fig:dSigmas}
	\end{figure} 

	\emph{The energy cost rate for faster speed of operation is minimized in the RC limit.---}
	 Lastly, we consider energy dissipation in the limit of the RC cycle. While $\QQ$ in general provides an efficiency measure of dissipation and product output precision, it is independent of time and hence agnostic to the cost of driving a cycle up to a required speed of operation. In this regard, Mallory et al.~\cite{KineticProofMallory} have shown that the ribosome and T7 polymerase prioritize speed over dissipation, and is therefore interesting to see how dissipation and mean production time vary between physiological systems and their corresponding RC limits. In particular, we calculate the difference in energy dissipation between the actual systems and their RC limits.
	 Figure \ref{fig:dSigmas} shows the normalized dissipation rate difference, $\dot{q}-\dot{q}_{\mathsmaller{\mathrm{RC}}} \equiv \kB T (\sigma - \sigma_{\mathsmaller{\mathrm{RC}}})/\Delta\mu_p$,
	 against the mean production time $\tau$ as a parametric function of $k$ for ribosomes and the T7 polymerase. The dissipation rate was normalized by the operating energy cost $\Delta\mu_p$ to allow comparison of different reaction networks. Evidently, while the ribosomes display absolute differences lower than the polymerase, they operate more slowly by two orders of magnitude and with steep energy costs for $\tau$ shorter than physiological values. On the other hand, the T7 polymerase maintains a relatively flat profile over many $\tau$ decades, ensuring that the energy dissipation rate does not deviate strongly from the ideal RC values, which achieve minimal $\QQ$ by construction.
	
	In closing, by all metrics considered, operating near the the RC limit confers considerable performance advantages to the KPR systems examined. By this measure, it is not surprising the polymerase outperforms the ribosomes given that its binding discrimination factor $f_1$ is about a million times more restrictive than that of the ribosomes ($f_{1,\mathrm{polymerase}}/f_{1,\mathrm{ribosome}} \sim 10^{-6}$) and places it significantly closer to the underlying RC limit. Note that a low $\QQ$ score \emph{does not} imply by itself the RC limit; low values of $\QQ$ are achieved for certain limiting values of $f_3$, and as discussed previously, this does not confer the similar trade-off advantages of approaching the RC limit. For instance, the Acc mutant achieves lower $\QQ$ score due to its enhanced $f_3$, but must operate at slower production times than the WT due to steeply increasing energy demands as seen in Figure \ref{fig:dSigmas}. As a result, operating near the RC limit not only achieves low $\QQ/\QQ_{\mathrm{lh}}$ score by definition, but also improves the overall \emph{global} performance per production cycle in a reaction network given a fixed energy budget.  

\section{Conclusions}
\label{sec:Conclusion}
	The ribosome and the DNA polymerase drive two essential production networks in the cell. The efficiency of these circuits is an important determinant of the organism fitness, and therefore they must be tuned to prioritize product-forming transitions over competing incorrect substrate binding and proofreading cycles. In this work, we have analyzed these circuits in the light of the Thermodynamic Uncertainty Relation (TUR), and found that the TUR measure $\QQ$ for the product current is closer to the lower bound in the polymerase than in the E. coli ribosome system. In particular, we considered a reduced cycle (RC) limit that accounts for paths leading to product formation, and showed that operating near this regime affords minimized values of $\QQ$ for corresponding rate constants. Notably, the polymerase operates very near the RC regime and thereby achieves nearly-optimal performance manifested by the proximity of its $\QQ$ measure to the lower bound $\QQ_\mathrm{lh}$ (Eq. \ref{eq:Qbound} and Table \ref{tab:Qscores}). Further, operating near RC relaxes the trade-off constraint between accuracy and speed, while decoupling both these measures from $\QQ$. On the other hand, a similar analysis showed that E. coli ribosomes operate relatively farther away from the RC limit, resulting in stronger coupling across all performance measures and increased energy costs, manifested by larger values of $\QQ$. 
	That said, the ribosome is not more than one order-of-magnitude away from the TUR bound. The significant difference in the performance of polymerase and ribosome stems from the accuracy of substrate discrimination, which is higher by about six orders of magnitude in the polymerase.
	This binding selectivity difference, which is not directly addressed here, is linked to the different biochemical mechanisms employed by the polymerase and the ribosome~\cite{KineticPoly1,KineticRibo2,RibosomeDecoder}. As a result, the polymerase is more likely to operate in the regime of correct product cycles than the ribosome, close to the RC limit. The different regimes of performance may also reflect the much more deleterious impact of errors in replication, which are carried through genome heredity, relative to errors in translation that vanish when the protein is degraded. For future studies, it would be interesting to study how distinct reaction pathways in other protein systems, \eg in signal transduction, adjust to prioritize correct response cycles and whether these imply similar RC limits that optimize the underlying TUR constraint.

\begin{acknowledgments}
The authors thank Anatoly B. Kolomeisky, Oleg A. Igoshin and Changbong Hyeon for helpful discussions.
This work was supported by the taxpayers of South Korea through the Institute for Basic Science, Project Code IBS-R020-D1.
\end{acknowledgments}

\bibliographystyle{apsrev4-1.bst}
\bibliography{sources}
\end{document}